\documentclass[twocolumn]{aastex631} 

\usepackage{cleveref}
\usepackage{textcomp}
\usepackage{verbatim}
\usepackage[T1]{fontenc}
\usepackage[scaled=0.85]{beramono}
\usepackage{animate}
\usepackage{natbib}

\definecolor{blue1}{RGB}{51, 153, 255} 
\definecolor{pink1}{RGB}{255, 102, 178}
\definecolor{pink2}{RGB}{255, 153, 255} 
\definecolor{purple1}{RGB}{204, 153, 255} 
\definecolor{purple2}{RGB}{153, 51, 255} 
\definecolor{blue1}{RGB}{51, 153, 255} 

\newcommand{\Ha}{H$\alpha$}

\newcommand{\OIIIHb}{[OIII] + H$\beta$}

\begin{document}

\title{Spectroscopy from Photometry: A Population of Extreme Emission Line Galaxies at $1.7 \lesssim z \lesssim 6.7$ Selected with JWST Medium Band Filters}

\author[0009-0000-8716-7695]{Sunna Withers}
\affiliation{Department of Physics and Astronomy, York University, 4700 Keele St. Toronto, Ontario, M3J 1P3, Canada\\}

\author[0000-0002-9330-9108]{Adam Muzzin} 
\affiliation{Department of Physics and Astronomy, York University, 4700 Keele St. Toronto, Ontario, M3J 1P3, Canada\\}

\author[0000-0002-5269-6527]{Swara Ravindranath}
\affiliation{Space Telescope Science Institute (STScI), 3700 San Martin Drive, Baltimore, MD 21218, USA\\}

\author[0000-0001-8830-2166]{Ghassan T. Sarrouh}
\affiliation{Department of Physics and Astronomy, York University, 4700 Keele St. Toronto, Ontario, M3J 1P3, Canada\\}

\author[0000-0002-4542-921X]{Roberto Abraham}
\affiliation{David A. Dunlap Department of Astronomy and Astrophysics, University of Toronto, 50 St. George Street, Toronto, Ontario, M5S 3H4, Canada\\}

\author[0000-0003-3983-5438]{Yoshihisa Asada}
\affiliation{Department of Astronomy and Physics and Institute for Computational Astrophysics, Saint Mary's University, 923 Robie Street, Halifax, Nova Scotia B3H 3C3, Canada\\}
\affiliation{Department of Astronomy, Kyoto University, Sakyo-ku, Kyoto 606-8502, Japan\\}

\author[0000-0001-5984-0395]{Maru{\v s}a Brada{\v c}} 
\affiliation{Department of Mathematics and Physics, Jadranska ulica 19, SI-1000 Ljubljana, Slovenia\\}

\author[0000-0003-2680-005X]{Gabriel Brammer}
\affiliation{Cosmic Dawn Center (DAWN), Denmark\\}
\affiliation{Niels Bohr Institute, University of Copenhagen, Jagtvej 128, DK-2200 Copenhagen N, Denmark\\}

\author[0000-0001-8325-1742]{Guillaume Desprez}
\affiliation{Department of Astronomy and Physics and Institute for Computational Astrophysics, Saint Mary's University, 923 Robie Street, Halifax, Nova Scotia B3H 3C3, Canada\\}

\author[0000-0001-9298-3523]{Kartheik Iyer}
\affiliation{Dunlap Institute for Astronomy and Astrophysics, 50 St. George Street, Toronto, Ontario, M5S 3H4, Canada\\}
\affiliation{Columbia Astrophysics Laboratory, Columbia University, 550 West 120th Street, New York, NY 10027, USA\\}

\author[0000-0003-3243-9969]{Nicholas Martis}
\affiliation{Department of Astronomy and Physics and Institute for Computational Astrophysics, Saint Mary's University, 923 Robie Street, Halifax, Nova Scotia B3H 3C3, Canada\\}
\affiliation{National Research Council of Canada, Herzberg Astronomy \& Astrophysics Research Centre, 5071 West Saanich Road, Victoria, BC, V9E 2E7, Canada\\}

\author[0000-0002-8530-9765]{Lamiya Mowla}
\affiliation{Dunlap Institute for Astronomy and Astrophysics, 50 St. George Street, Toronto, Ontario, M5S 3H4, Canada\\}

\author{Ga\"el Noirot}
\affiliation{Department of Astronomy and Physics and Institute for Computational Astrophysics, Saint Mary's University, 923 Robie Street, Halifax, Nova Scotia B3H 3C3, Canada\\}

\author[0000-0002-7712-7857]{Marcin Sawicki}
\affiliation{Department of Astronomy and Physics and Institute for Computational Astrophysics, Saint Mary's University, 923 Robie Street, Halifax, Nova Scotia B3H 3C3, Canada\\}

\author[0000-0002-6338-7295]{Victoria Strait}
\affiliation{Cosmic Dawn Center (DAWN), Denmark\\}
\affiliation{Niels Bohr Institute, University of Copenhagen, Jagtvej 128, DK-2200 Copenhagen N, Denmark\\}

\author[0000-0002-4201-7367]{Chris J. Willott}
\affiliation{National Research Council of Canada, Herzberg Astronomy \& Astrophysics Research Centre, 5071 West Saanich Road, Victoria, BC, V9E 2E7, Canada\\}







\begin{abstract}
We use JWST/NIRCam medium band photometry in a single pointing of the CAnadian NIRISS Unbiased Cluster Survey (CANUCS) to identify 118 Extreme Emission Line Galaxies (EELGs) over $1.7 \lesssim z \lesssim 6.7$, selected using a set of color cuts that target galaxies with extreme \OIIIHb\ and \Ha\ emission. We show that our medium band color selections are able to select galaxies based on emission line EW, which is advantageous to more commonly used selections since it does not require strong continuum emission, and can select galaxies with faint or red continuum fluxes. The median EWs of our sample is $EW(\text{H}\alpha) = 893 $ \AA\ and $ EW(\text{[OIII] + H}\beta) = 1255 $ \AA, and includes some objects with $EW(\text{[OIII] + H}\beta) \sim 3000$ \AA. These systems are mostly compact with low stellar mass (median $\log(M_\star/M_\odot) = 8.03$), low metallicity (median $Z = 0.14 Z_\odot$), little dust (median $A_V = 0.18$ mag) and high SSFR (median $SSFR = 1.18 \times 10^{-8}/yr$). Additionally, galaxies in our sample show increasing EW(\Ha) and EW(\OIIIHb) with redshift, an anti-correlation of EW(\Ha) with stellar mass, and no correlation between EW(\OIIIHb) and stellar mass. Finally, we present NIRSpec spectroscopy of 15 of the EELGs in our sample. These spectra confirm the redshifts and EWs of the EELGs calculated from the medium bands, which demonstrates the accuracy and efficiency of our color selections. Overall, we show that there are significant advantages to using medium band photometry to identify and study EELGs at a wide range of redshifts.
\end{abstract}


\section{Introduction} \label{introduction}

Extreme emission line galaxies (EELGs) have become a major field of study in recent years, with many searches targeting EELGs in the local Universe up to high redshifts (e.g. local Universe: \citealt{cardamone_galaxy_2009}, \citealt{henry_close_2018}, \citealt{liu_strong_2022}; intermediate-$z$: \citealt{van_der_wel_extreme_2011}, \citealt{maseda_nature_2014}, \citealt{maseda_number_2018}, \citealt{tang_mmtmmirs_2019}, \citealt{onodera_broadband_2020}, \citealt{tran_mosel_2020}, \citealt{boyett_early_2022}, \citealt{gupta_mosel_2023}; high-$z$: \citealt{stark_keck_2013}, \citealt{smit_evidence_2014}, \citealt{endsley_o_2021}, \citealt{kashino_eiger_2022}, \citealt{matthee_eiger_2022}, \citealt{asada_jwst_2022}, \citealt{williams_jems_2023}, \citealt{rinaldi_strong_2023}). Regardless of their redshift, EELGs are characterized by high equivalent width (EW) UV-optical emission lines driven by elevated star formation, generally found in low mass, metal-poor galaxies with little dust. While there is no universal definition, systems with emission line EW $\gtrsim 100$\AA\ are typically considered EELGs. However, this definition varies widely and is often dependent on redshift.

It is currently presumed that strongly star-forming galaxies are the main drivers of Hydrogen reionization over $5.5 \lesssim z \lesssim 15$ (e.g. \citealt{robertson_cosmic_2015}, \citealt{stefanon_high_2022}), which serves to motivate many of the searches for high-$z$ EELGs. While not all star-forming galaxies can be classified as EELGs, EELGs are abundant at high-$z$ (\citealt{stark_keck_2013}, \citealt{smit_evidence_2014}) and may play an important role during the epoch of reionization (EoR). With the beginning of JWST's science operations, it is now possible to more fully characterize of the population of EELGs up to $z \sim 9$ with NIRCam and NIRSpec. This includes studies of EELGs in the EoR, as well as at intermediate-$z$ ($z \lesssim 5$). These lower-$z$ EELGs can serve as analogues to high-$z$ star forming galaxies such as EELGs, and can provide valuable insights on the high-$z$ population, even in the era of JWST (e.g. \citealt{tang_stellar_2022}, \citealt{rhoads_finding_2023}, \citealt{mingozzi_classy_2022}).

Early JWST studies of EELGs have already begun to yield interesting results. These include the identification of very high EW systems (EW $> 1000$\AA\ for \OIIIHb\ and \Ha) up to $z \sim 9$, and confirmation that EELGs are common in the high-$z$ Universe (e.g. \citealt{matthee_eiger_2022}, \citealt{kashino_eiger_2022}, \citealt{williams_jems_2023}, \citealt{rinaldi_strong_2023}). Other work has shown that the Hydrogen ionizing photon production efficiencies ($\xi_{\text{ion}}$) of high-$z$ EELGs are generally high, as expected if these objects are responsible for reionization (e.g. \citealt{sun_first_2022}, \citealt{asada_jwst_2022}). Additionally, there have been several spectroscopic studies on the physical conditions of the interstellar medium, which find line ratios similar to what is seen in at intermediate redshifts, and low metallicities (e.g. \citealt{trump_physical_2022}, \citealt{taylor_metallicities_2022}). 

Many of the searches for EELGs select samples using either broad-band photometry (e.g. \citealt{stark_keck_2013}, \citealt{onodera_broadband_2020}) or wide field slitless spectroscopy (WFSS, e.g. \citealt{maseda_nature_2014}, \citealt{maseda_number_2018}, \citealt{kashino_eiger_2022} (EIGER), \citealt{boyett_early_2022} (GLASS)). However, while less common, medium-band photometry can provide a powerful tool in the search for EELGs (e.g. \citealt{terao_selection_2022}, \citealt{gupta_mosel_2023}, \citealt{williams_jems_2023}). The medium bands provide a finer wavelength sampling of galaxy spectral energy distributions (SEDs) than the broad bands, thus offering improved estimates of galaxy properties (e.g. \citealt{roberts-borsani_improving_2021}, Sarrouh et al. in preparation). Additionally, medium band imaging is free of many of the challenges associated with WFSS observations (such as overlapping source contamination), and can reach greater depths per unit exposure time than WFSS.  The medium bands can thus be more efficient at identifying and studying EELGs than more conventional methods.

A simple and effective way of searching for EELGs using medium band photometry is by using color selections. These color selections target the extreme colors produced by extreme emission lines, which can reach medium band colors $> 2$ mag in neighbouring filters (e.g. \citealt{williams_jems_2023}). One advantage of medium band color selections is that they can select galaxies based on emission line flux. Unlike the commonly used Lyman break technique, medium band color selections do not require high signal-to-noise ratios (S/N) in the rest-frame UV-optical continuum, and can select EELGs which are too faint to be detected through their continuum emission alone. This population of very faint galaxies may play an important role in reionization (e.g. \citealt{endsley_jwstnircam_2022}), yet will not be detected using many of the typical selection criteria. However, faint and low-mass EELGs often have strong emission lines, allowing them to be selected using medium-band color cuts. Additionally, EELGs with red rest-frame UV-optical continuum emission have been shown to mimic the signatures of very high-$z$ Lyman break galaxies selected through their broad-band photometry (e.g. CEERS-93316, \citealt{arrabal_haro_spectroscopic_2023}). We will show that objects such as these can be easily identified using medium band color selections, and can thus be used to identify lower-$z$ contaminants in samples with medium band imaging.




\begin{figure*}
    \centering
    
    \animategraphics[width = 0.85\linewidth, autoplay, loop, controls]{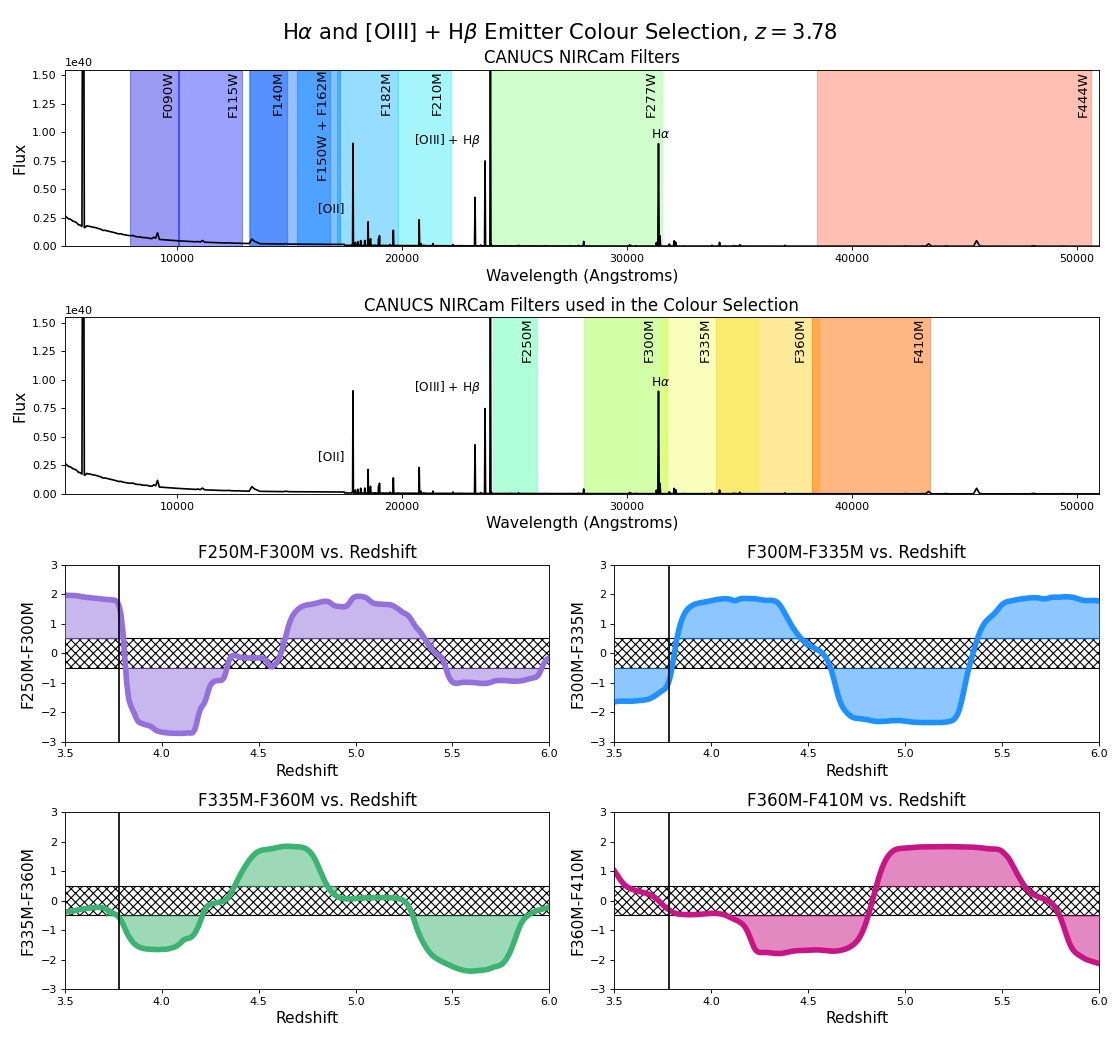}{}{1}{125}
    
    \caption{An animated demonstration of the color cuts used in this work. The top two panels show a \texttt{Yggdrasil} SED overlaid on filter the coverage for the filters used in the example color selection (second panel, F250M, F300M, F335M, F360M, and F410M) and the rest of the filters used in the CANUCS NIRCam flanking fields (first panel, F090W, F115W, F140M, F150W, F162M, F182M, F210M, F277W, and F444W). The bottom four panels show how four colors evolve with galaxy redshift, which are used to select EELGs over $3.5 \lesssim z \lesssim 5.5$. The hatched regions indicate colors of $\pm 0.5$, demonstrating how the \Ha\ and \OIIIHb\ emission lines drive strong color excesses in the medium bands. As an example of the color cuts, these medium band filters can be used to select EELGs at $ 4.8 \lesssim z \lesssim 5.5$ by selecting galaxies with two extreme colors: $F360M - F410M > 0.5$ driven by strong \Ha\ emission and $F300M - F335M < -1$ driven by strong \OIIIHb\ emission. The selection criteria shown here is just an example, and the full set of color cuts will be presented in a future work. Adobe Reader is required to view the animation, alternatively, it is available to view \href{https://www.youtube.com/watch?v=bTjjjyFRJys}{here} and available for download \href{https://github.com/sunnawithers/EELG_colours/blob/main/Figure1_EELGcolours_Withers\%2B2023.mp4}{here}. }
    \label{fig:fig1}
\end{figure*}

This work presents a sample of 118 EELGs over $1.7 \lesssim z \lesssim 6.7$. These objects were identified using a set of NIRCam medium band color cuts that target galaxies with extreme \OIIIHb\ and \Ha\ emission, described in \textsection \ref{sec:Observations and sample selection}. We determine the \OIIIHb\ and \Ha\ EWs (\textsection \ref{sec:EWs}) and physical properties (\textsection \ref{sec:physical properties}) for the sample, and study the evolution of EWs with physical properties of the galaxy and redshift (\textsection \ref{sec:properties dependence}). Additionally, we present follow-up spectroscopy of 15 of the objects with NIRSpec, which we will use to demonstrate the ability of the medium bands to obtain robust measurements of EW(\Ha) and EW(\OIIIHb) (\textsection \ref{sec:SPEEEEECCCTTTRRRRRRAAAAAAAAAA}). We assume a cosmology of $H_0 = 72$km s$^{-1}$ Mpc$^{-1}$, $\Omega_m = 0.27$, and $\Omega_\Lambda = 0.73$. 

\begin{figure*}
    \centering
    \includegraphics[width = 0.8\textwidth]{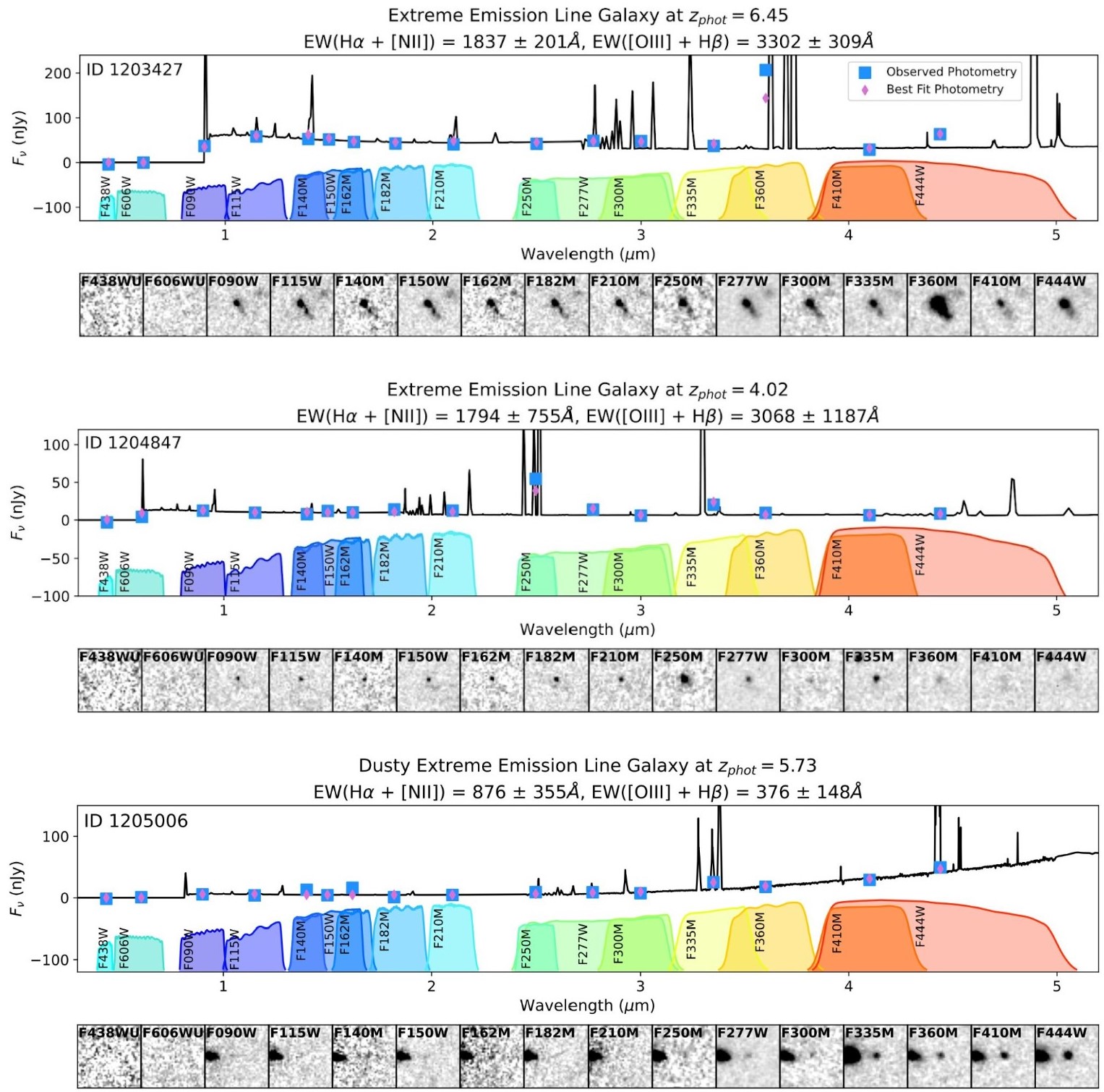}
    \caption{Examples of three EELGs identified in this work. Each example has $3'' \times 3''$ cutouts of the galaxy in the fourteen NIRCam filters and two WFC3/UVIS filters used in this work (bottom panels), as well as the filter throughput curves, photometry (blue squares), the best fit SED from \texttt{EAZY} (black curves), and the predicted photometry from \texttt{EAZY} (purple diamonds). The first two SEDs (ID 1203427 and 1204847) have blue continuum fluxes and strong Lyman breaks, however the third example (ID 1205006) has a very red continuum and no visible Lyman break. This example demonstrates the ability of the medium bands to select EELGs without the need for strong continuum emission, which can select faint galaxies and identify objects such as CEERS-93316 (\citealt{arrabal_haro_spectroscopic_2023}). The best fit SEDs (black curves) and predicted photometry (purple diamonds) in this figure are from photometric redshift code \texttt{EAZY} (\citealt{brammer_eazy_2008}), using the set of templates outlined in \citealt{larson_spectral_2022}. The best-fit SEDs were generated by forcing \texttt{EAZY} to fit the photometry at the redshift implied by the color selections.}
    \label{fig:fig2}
\end{figure*}

\section{Observations \& Sample Selection} \label{sec:Observations and sample selection}

\subsection{Observations} \label{sec:observations}

This work uses NIRCam imaging obtained as part of the CAnadian NIRISS Unbiased Cluster Survey (CANUCS, \citealt{willott_near-infrared_2022}, GTO ID 1208). CANUCS observations include deep ($\sim 28.9$ mag in medium bands, $\sim 29.4$ mag in wide bands to $5 \sigma$ for point sources) NIRCam imaging of flanking fields in five broad (F090W, F115W, F150W, F277W, F444W) and nine medium band (F140M, F162M, F182M, F210M, F250M, F300M, F335M, F360M, F410M) filters. This work uses the first of five CANUCS NIRCam flanking fields near the cluster MACS J0417.5-1154. The data were obtained in a single NIRCam pointing using modules A and B covering an area of $9.7$ arcmin$^2$. Additionally, $\sim 70$\% of the flanking field was observed in two HST WFC3/UVIS filters (F438W and F606W) to depths of $\sim 28.4$ mag at $5\sigma$ for point sources (Program ID 16667, PI M. Bradac). Together, these observations provide a wavelength coverage of $\lambda \sim 0.4 - 5 \mu$m. Lensing magnification from the cluster is negligible in the flanking field, and is thus neglected from this analysis.

The NIRCam data were processed using a combination of the official STScI JWST pipeline (with software version 1.8.0 and CRDS context \texttt{jwst\_1001.pmap}) and \texttt{grizli} version 1.6.0 (\citealt{brammer_gbrammergrizli_2021}). The NIRCam and HST data were drizzled onto the same pixel scale of 40 milliarcsec/pixel, and registered to Gaia DR3 astrometry. Further details of our processing method are provided in \cite{noirot_first_2022}. Empirical point spread functions (PSFs) were then built from isolated stars in the field and all images were convolved with kernels to match the PSF of the longest wavelength filter, F444W. These PSF-matched images are used for photometry to ensure colors are not biased by differences in PSF. 

\begin{figure*} 
    \begin{center}
        \includegraphics[width = 0.9\textwidth]{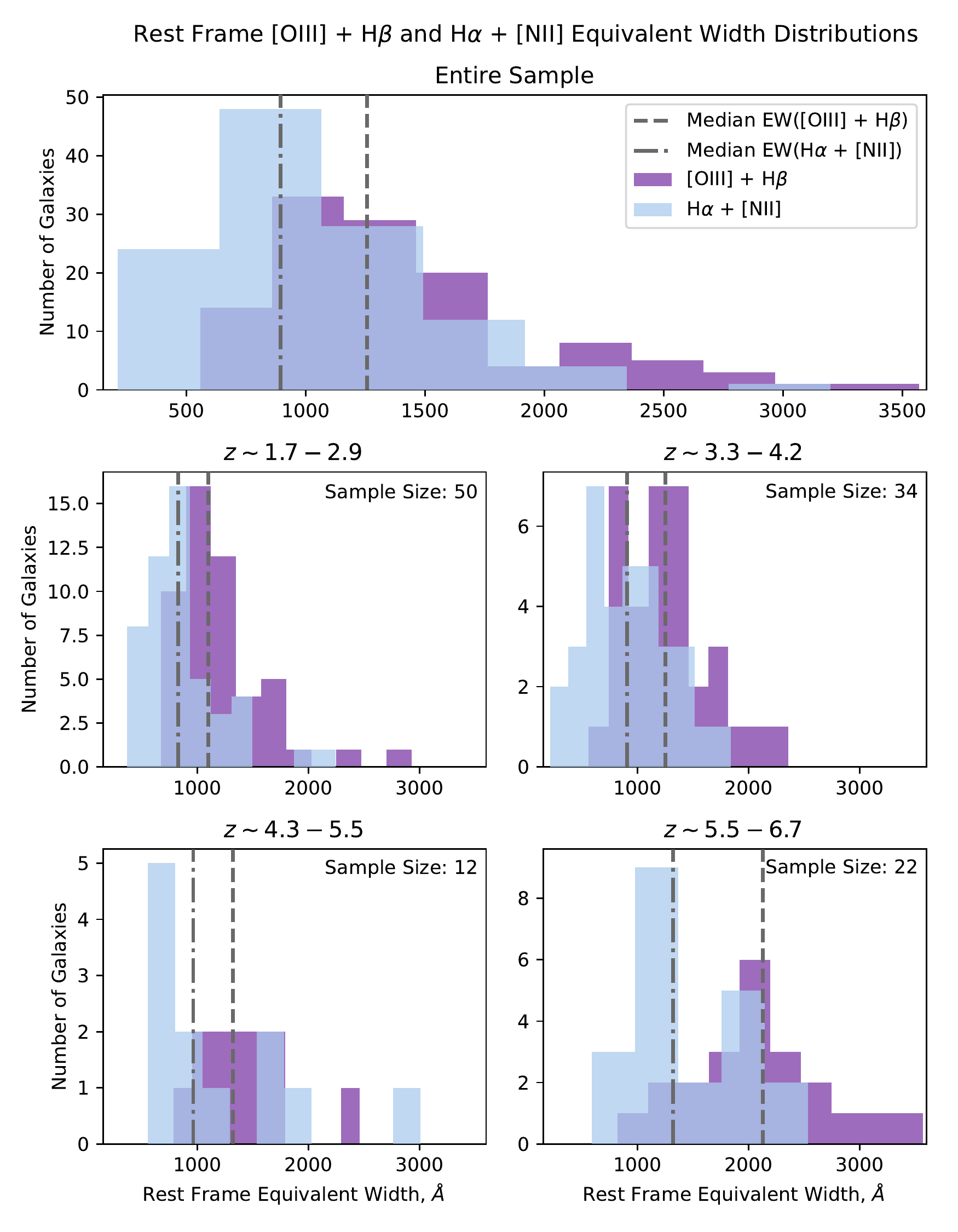}
    \end{center}
    \caption{Histograms showing EW(\OIIIHb) (purple) and EW(\Ha) (blue) for the entire sample (top panel) and as a function of redshift (bottom four panels). Each panel also shows the median EW(\OIIIHb) (gray dashed line) and median EW(\Ha) (gray dash dotted line). This figure demonstrates that very strong EELGs do exist over $z\sim 1.7 - 6.7$, reaching EW $> 1000$\AA, and that the median EW(\Ha) and EW(\OIIIHb) increases with redshift.}

    \label{fig:fig3}
\end{figure*}

\subsection{Color Selections} \label{sec:color selections}

Our sample (see \textsection \ref{sec:sample}) was selected using a set of color cuts that targets galaxies with both extreme \OIIIHb\ and \Ha\ emission. The color cuts were defined by creating synthetic NIRCam observations of SEDs produced by the \texttt{Yggdrasil} stellar population synthesis code (\citealt{zackrisson_spectral_2011}). In order to define our color selections, we selected a subset of the publicly available \texttt{Yggdrasil} SEDs with a \cite{kroupa_variation_2001} initial mass function (IMF), two star formation histories (instantaneous burst and constant), with various ages ($0.01 - 9.1$ Myr after star formation began), and metallicities ($Z = 0.0004, 0.004, 0.008$). The magnitudes in the 16 CANUCS filters were then calculated for each SED redshifted to $z = 0.1 - 15$ (with $\Delta z = 0.1$ step sizes). Additionally, we accounted for the influence of interloper galaxy populations (such as old and dusty galaxies) by creating synthetic observations of SEDs from \texttt{Flexible Stellar Population Synthesis (FSPS)} code (\citealt{conroy_propagation_2009}, \citealt{conroy_propagation_2010}). We used this set of synthetic NIRCam observations to search for medium band color excesses driven by strong \OIIIHb\ and \Ha\ emission, which we use to define an initial set of color cuts. With this initial set, we analyse how effective each color cut is by evaluating the completeness and contamination fractions when employing different S/N cuts and colors (e.g., selections requiring $0.5$ mag color vs. $1$ mag color). 

With the results of these tests, we define a set of color cuts that select galaxies with extreme \OIIIHb\ and \Ha\ emission over $1.7 \lesssim z \lesssim 6.7$. The tests described above show that our color selections can effectively identify galaxies with EW$(\text{H}\alpha) \gtrsim 500 $ \AA\ and  EW$ (\text{[OIII] + H}\beta) \gtrsim 1000 $ \AA\ at all redshifts targeted in this work. As discussed in \textsection1, medium band color cuts are able to select EELGs based on line flux without the need for strong continuum emission. Nonetheless, we employ a S/N cut of S/N $\geq 4$ in line emission and S/N $\geq 2$ on average in the underlying continuum. We choose to implement this S/N cut since our tests show that it reduces the number of contaminants in our sample, however this selection will be expanded upon in future work. Additionally, while it is possible to identify EELGs by targeting emission from a single emission line complex, our color selections require strong emission in both \OIIIHb\ and \Ha. Much like the S/N cuts, we choose to do this since our tests show that selecting galaxies on two emission lines provides a more reliable estimate of galaxy redshift than with one line.

The animation in Figure \ref{fig:fig1} illustrates the color cuts used in this work, in which the colors of five neighbouring medium bands are used to target galaxies over $3.5 \lesssim z \lesssim 5.5$. As an example, one of these color cuts selects galaxies over $4.8 \lesssim z \lesssim 5.5$ by targeting galaxies with two extreme colors: $F360M - F410M > 0.5$ driven by strong \Ha\ emission and $F300M - F335M < -1$ driven by strong \OIIIHb\ emission. Overall, our selection criteria consists of ten color cuts which were chosen based on how effective they were at selecting EELGs at the correct redshifts. The full set of ten color cuts will be presented in a future work, along with further discussion on how the color selections were defined.

\subsection{Sample} \label{sec:sample}

Based on the color and S/N cuts described above, we select 118 objects over $1.7 \lesssim z \lesssim 6.7$ with extreme \OIIIHb\ and \Ha\ emission. As noted in \textsection1, color cuts similar to those used in this work can select galaxies based on emission-line EW, with no dependence on strong rest frame UV-optical continuum. This is advantageous to more typically used selections for high-$z$ galaxies, since it is able to target galaxies with faint or red continuum fluxes for which a Lyman break cannot be observed. This is exemplified in Figure \ref{fig:fig2}, which shows three examples of EELGs found using the medium band color selections. All three examples show flux excesses in two filters driven by strong \OIIIHb\ and \Ha\ emission. Two of the examples (ID 1203427 and 1204847) exhibit blue continuum slopes and show clear evidence of a Lyman break. However, the third example (ID 1205006) has a very red continuum, which results in an absence of a Lyman break. Objects such as these are similar to CEERS-93316 (\citealt{arrabal_haro_spectroscopic_2023}), which have been shown to contaminate samples of photometrically selected very high-$z$ galaxies (as discussed in \textsection1). The ability of medium band color cuts to select galaxies such as these can provide a powerful tool in the search for very high-$z$ galaxies by identifying lower-$z$ interlopers in those samples.

    
\begin{figure*}[]
    \centering
    \includegraphics[width = 0.9\textwidth]{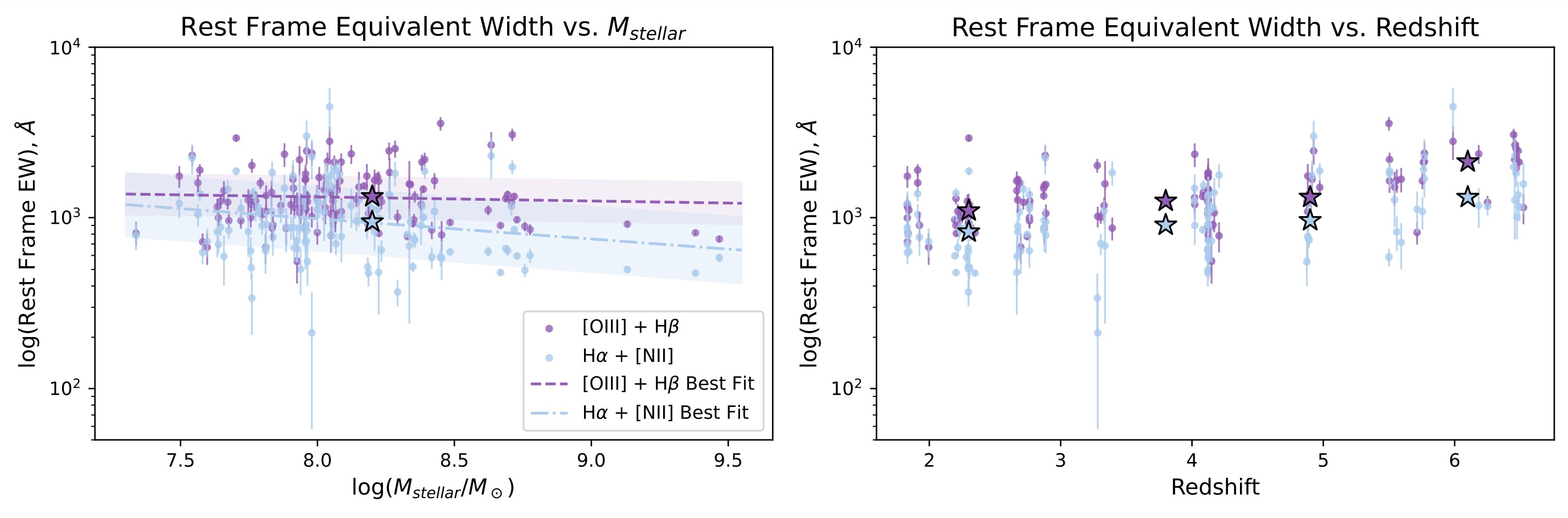}
    \caption{Left: EW(\OIIIHb) (purple) and EW(\Ha) (blue) as a function of stellar mass, with a line of best fit for EW(\OIIIHb) (dashed line) and EW(\Ha) (dash dot line). The median EWs for galaxies with $7.9 < \log(M_\star / M_\odot) < 8.5$ are plotted as blue (\Ha) and purple (\OIIIHb) stars. Galaxies in this sample reproduce the known anti-correlation between EW(\Ha) and stellar mass, but show no evolution of EW(\OIIIHb) and stellar mass. Right: EW(\OIIIHb) (purple) and EW(\Ha) (blue) as a function of redshift. Also plotted is the median EWs for galaxies in the same four redshift bins as Figure \ref{fig:fig3} with $7.9 < \log(M_\star / M_\odot) < 8.5$ as the blue (\Ha) and purple (\OIIIHb) stars. This panel shows how the EWs of galaxies in our sample increases with redshift, which is likely driven by increasing SSFRs at earlier times. }
    \label{fig:fig4}
\end{figure*}

\section{Analysis} \label{sec:Analysis}

\subsection{Rest Frame Equivalent Widths} \label{sec:EWs}

Rest frame EW(\OIIIHb) and EW(\Ha) were calculated directly from the photometry by fitting the continuum flux to a power law, $f_\nu \propto \lambda^\alpha$. Different continuum filters were used for each of the ten selection criteria, and were chosen by avoiding filters that may be contaminated with emission lines (such as [OII] or Pa$\beta$) at the galaxy redshift. Photometric redshifts were determined using \texttt{EAZY} (\citealt{brammer_eazy_2008}), with templates from \texttt{FSPS} (\citealt{conroy_propagation_2009}, \citealt{conroy_propagation_2010}) and \cite{larson_spectral_2022}, by limiting the allowed best fit redshifts to those implied by the color selections. This allowed us to obtain a precise redshift for each galaxy that is consistent with the findings of the color selections. The EWs calculated this way will be overestimated, since the contribution from weaker emission lines (such as [NII] for \Ha\ and the Balmer lines for \OIIIHb) was not removed prior to calculations. However, it is not expected that these lines will contribute significantly to the EW compared to the much stronger \OIIIHb\ and \Ha\ emission lines (e.g. \citealt{izotov_low-redshift_2019}, \citealt{cameron_jades_2023}). 

Figure \ref{fig:fig3} shows the distribution of EW(\OIIIHb) and EW(\Ha) for the entire sample (top panel), and with the sample separated into four redshift bins (bottom panels). This figure demonstrates that very high EW systems do exist over $1.7 \lesssim z \lesssim 6.7$, with EW $> 1000$\AA\ in both \OIIIHb\ and \Ha, sometimes even reaching EW $\sim 3000$\AA. Other searches for EELGs have revealed similarly high EW objects (e.g. \citealt{endsley_o_2021}, \citealt{stefanon_high_2022}, \citealt{rinaldi_strong_2023} at high-$z$, \citealt{reddy_mosdef_2018}, \citealt{boyett_early_2022}, \citealt{onodera_broadband_2020} at intermediate-$z$), however, the most extreme objects are quite rare and represent only a small sample of the total star forming galaxies at any given redshift.

\subsection{Physical Properties} \label{sec:physical properties}

The \texttt{Dense Basis} (\citealt{iyer_nonparametric_2019}) SED fitting code was used to compute the physical properties of EELGs in the sample. Fitting was performed using Kron aperture photometry and assuming the \cite{calzetti_dust_2001} dust law, \cite{kroupa_variation_2001} IMF, and at the best fit redshift from \texttt{EAZY} (see \textsection \ref{sec:EWs}). Additionally, \texttt{Dense Basis} was allowed stellar masses of $ 7 < \log(M_\star/M_\odot) < 12$, dust attenuation of $0 < A_V < 4$, and metallicities of $-2 < \log(Z/Z_\odot) < 0.3$. The \texttt{Dense Basis} fitting shows that galaxies in our sample are typically low mass (median $\log(M_\star/M_\odot) = 8.02 $), low metallicity (median $Z = 0.14 Z_\odot$), with little dust attenuation (median $A_V = 0.18 $ mag), and high SSFR (median $SSFR = 1.18 \times 10^{-8}/yr$). Furthermore, a visual inspection of the images reveals most of the EELGs are compact with little to no structure. However, there is a small number of EELGs that show evidence of interactions or mergers, such as example 1 (ID 1203427) of Figure \ref{fig:fig2}. Other works on EELGs find similar properties, at both high-$z$ (e.g. \citealt{sun_first_2022}, \citealt{asada_jwst_2022}) and intermediate-$z$ (e.g. \citealt{tang_mmtmmirs_2019}). 

\subsection{Dependence of Equivalent Widths on Physical Properties and Redshift}   \label{sec:properties dependence}

Figure \ref{fig:fig4} shows the dependence of rest frame \OIIIHb\ and \Ha\ EWs with stellar mass (left) and redshift (right), including the median values for galaxies with $ 7.9 < \log(M_\star/M_\odot) < 8.5$ (a mass range that is covered in each of the redshift bins used in Figure \ref{fig:fig3}). The left and right panels of Figure \ref{fig:fig4} illustrate several important points about our sample of EELGs. First, a visual inspection of the left panel of Figure \ref{fig:fig4} shows that our sample reproduces the known anti-correlation between rest frame EW(\Ha) and stellar mass (e.g. \citealt{matthee_eiger_2022}, \citealt{onodera_broadband_2020}, \citealt{rinaldi_strong_2023}). However, there does not appear to be any correlation between rest frame EW(\OIIIHb) and stellar mass. To test this, we performed the Spearman's rank-order correlation test on the distributions. The tests yield $t_{\text{H}\alpha} = -0.18$, $p_{\text{H}\alpha} = 0.045$, and $t_{\text{[OIII] + H}\beta} = -0.0033$, $p_{\text{[OIII] + H}\beta} = 0.97$, confirming that there exists an anti-correlation between EW(\Ha) and stellar mass, and no correlation of EW(\OIIIHb) and stellar mass. Taken at face value, this trend of constant EW(\OIIIHb) with stellar mass indicates that metallicity of the EELGs may remain roughly constant with stellar mass. However, this conclusion is only tentative given that our sample contains selection biases for the strongest EWs, and therefore may be incomplete at lower EWs.

Additionally, the right panel of Figure \ref{fig:fig4} shows evidence of increasing EW(\OIIIHb) and EW(\Ha) with redshift, a trend which is also observed in Figure \ref{fig:fig3}. The same behaviour has been observed in other populations of EELGs (e.g. \citealt{reddy_mosdef_2018}, \citealt{matthee_eiger_2022}), and is generally attributed to higher SSFRs at earlier times (e.g. \citealt{reddy_mosdef_2018}).

\section{Robustness of Estimating Equivalent Width's with the Medium Bands} \label{sec:SPEEEEECCCTTTRRRRRRAAAAAAAAAA}

We obtained follow-up spectroscopy for 15 of the objects presented in this paper, acquired with NIRSpec in multi-object spectroscopy (MOS) mode using $\sim$ 3ks integration times. Spectra were taken using the prism, which provides low-resolution spectroscopy ($R \sim 100$) over $\lambda \sim 0.6 - 5.3 \mu$m. The spectroscopy processing was performed with a combination of the official STScI JWST pipeline (with software version 1.8.4 and CRDS context \texttt{jwst\_1030.pmap}) and the \texttt{msaexp} package (\citealt{brammer_msaexp_2022}). Level 1 processing of raw data into count rate files used the standard pipeline with the jump step option \texttt{expand\_large\_events} enabled to mitigate snowball residuals, and a custom persistence correction that masks pixels that approach saturation for any read-out groups within the subsequent 1200\,s, based on an analysis of the typical NIRSpec detector persistence timescale. Level 2 processing performed the standard wavelength, flat-field, path-loss correction and photometric calibration steps. Individual 2D spectra were combined and optimal 1D extractions (\citealt{horne_optimal_1986}) were done using \texttt{msaexp}.

\begin{figure*}
    \centering
    \includegraphics[width = 1\textwidth]{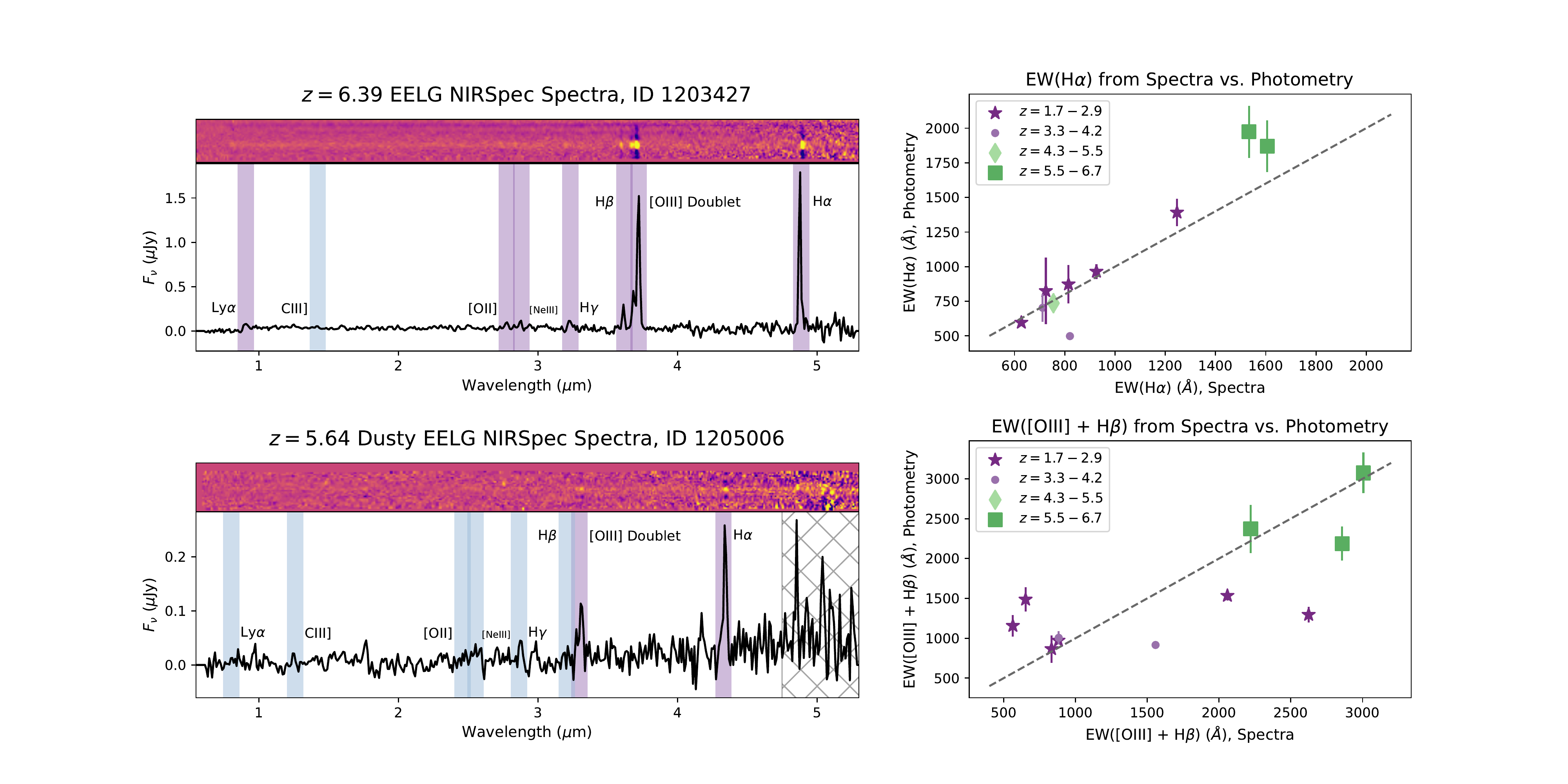}
    \caption{Left: Examples of the NIRSpec spectrum's of two EELGs in the sample. The top panel corresponds to the first example in Figure \ref{fig:fig2} (ID 1203427), and the bottom panel corresponds to the third example in Figure \ref{fig:fig2} (ID 1205006). Various optical emission lines are labelled on the spectra, in purple if they are detected and in blue if they are not. Strong \Ha\ and \OIIIHb\ emission is observed in both spectra, confirming that the color cuts are able to select EELGs and accurately determine their redshift. Right: EW(\Ha) (top) and EW(\OIIIHb) (bottom) calculated from the NIRSpec spectra vs. the medium band photometry. While there is some scatter in the plots, the EWs calculated from the spectra and photometry do generally agree with each other, however the agreement is better for EW(\Ha) than for EW(\OIIIHb). These differences are likely driven by uncertainties in continuum estimates when calculating EWs from the medium band photometry. That said, the right hand figures show that it is possible to accurately calculate EWs using medium band photometry, when averaged over a multi-object sample. }

    \label{fig:fig5}
\end{figure*}

The left panels of Figure \ref{fig:fig5} show two examples of our spectra, which correspond to the first (ID 1203427, top panel) and third (ID 1205006, bottom panel) examples in Figure \ref{fig:fig2}. The spectroscopic redshifts of all 15 spectra agree with those implied by the color selections, demonstrating the effectiveness of medium band color cuts at accurately selecting EELGs over $1.7 \lesssim z \lesssim 6.7$. Additionally, the right panels plot the EW(\Ha) and EW(\OIIIHb) calculated from photometry versus spectroscopy. The spectroscopic EWs were calculated by fitting the continuum surrounding the emission lines with a straight line, which was then used to estimate the underlying continuum emission. The continuum fluxes need to be accurately measured in order to calculate EWs, which requires sufficient S/N in the continuum. While the \Ha\ and \OIIIHb\ emission lines are well detected in all 15 spectra, several galaxies had continua that were too faint to reliably calculate EWs, and were thus excluded from this analysis. Additionally, we were unable to measure the \OIIIHb\ lines for two galaxies, since that portion of the spectrum fell in the NIRSpec detector gaps. In total, EW(\Ha) was calculated for ten spectra, and EW(\OIIIHb) was calculated for thirteen.

The right panels of Figure \ref{fig:fig5} show a general agreement between EWs calculated from medium band photometry and from the spectra, and confirm the existence of very high EW systems with EW(\OIIIHb) $\sim 3000$\AA. While the EWs calculated from medium band photometry and spectroscopy do generally agree, the agreement is not perfect and is most pronounced for EW(\OIIIHb). The main cause of this disagreement is likely driven by uncertainties in continuum estimates when EWs are calculated from the photometry, for two reasons. First, it can be difficult to obtain reliable continuum estimates, especially in the region surrounding the \OIIIHb\ line complex, as there exists significant contamination from other emission lines in this region. This causes the continuum to be poorly constrained resulting in incorrect EW measurements, particularly for lower-$z$ galaxies. Furthermore, the EWs were calculated assuming all of the flux from emission lines falls in one filter. However, this assumption does not always hold true and can lead to EWs being underestimated when calculated from medium band photometry. A more advanced technique for dealing with these issues is beyond the scope of this Letter, and will be investigated in detail in future work. However, the overall agreement between EWs from photometry and spectroscopy in Figure \ref{fig:fig5} is sufficient to state that in general, it is possible to accurately calculate EWs from medium band photometry.

\section{Conclusion}

This Letter presents an analysis of 118 extreme \Ha\ and \OIIIHb\ emitters over $1.6 \lesssim z \lesssim 6.7$ selected from NIRCam observations as part of the CANUCS GTO program. These sources were identified using a set of medium band color cuts, and include some very red galaxies (see the third example (ID 1205006) in Figure \ref{fig:fig2}), which have continuum fluxes rewards of the Lyman break that are too faint to be detected. Objects such as these demonstrate the ability of our color cuts to select galaxies on emission line flux, with no dependence on continuum emission. Thus, we show medium band color selections are a powerful tool that can identify very faint galaxies which may play an important role in reionization, and objects that contaminate photometrically selected samples of very high-$z$ galaxies.

We calculate the EWs (\textsection \ref{sec:EWs}) and physical properties (\textsection \ref{sec:physical properties}) of the EELGs in our sample. They all have very high \OIIIHb\ and \Ha\ EWs, sometimes up to $\sim 3000$\AA. Their physical properties are typical of EELGs: low mass, low metallicity, high SSFRs, and compact morphologies. Additionally, we show that our EELGs in our sample follow trends of increasing EW(\Ha) and EW(\OIIIHb) with redshift, an anti-correlation between EW(\Ha) and stellar mass, and no correlation between EW(\OIIIHb) and stellar mass (\textsection \ref{sec:EWs}, \ref{sec:properties dependence}, Figures \ref{fig:fig3}, \ref{fig:fig4}). Finally, we present follow-up NIRSpec spectroscopy of 15 of these sources (\textsection \ref{sec:SPEEEEECCCTTTRRRRRRAAAAAAAAAA}). These spectra confirm that medium band color cuts are able to efficiently select EELGs over a wide range of redshifts, and can be used to accurately calculate EWs when averaged over multi-object samples (Figure \ref{fig:fig5}). 

It is important to reiterate that the color selections used in this work target the brightest and most extreme EELGs with strong \Ha\ and \OIIIHb\ emission, in the first of five CANUCS fields. This leaves open many avenues for future work that will allow us to significantly expand the sample of galaxies presented and discussed in this Letter. These include searches for fainter and less extreme EELGs, color selections involving only one emission line complex, incorporation of full photometric redshift fitting with appropriate templates, and analysis of the remaining four CANUCS NIRCam flanking fields.

\begin{acknowledgments}

 This research was enabled by grant 18JWST-GTO1 from the Canadian Space Agency and funding from the Natural Sciences and Engineering Research Council of Canada.

\end{acknowledgments}

\bibliography{Citations_v6}{}
\bibliographystyle{aasjournal}



\end{document}